\documentclass[]{mn2e}
\renewcommand\[{\begin{equation}}
\renewcommand\]{\end{equation}}

\catcode`\@=11
\def\gsim{\ifmmode{\mathrel{\mathpalette\@versim>}}
    \else{$\mathrel{\mathpalette\@versim>}$}\fi}
\def\lsim{\ifmmode{\mathrel{\mathpalette\@versim<}}
    \else{$\mathrel{\mathpalette\@versim<}$}\fi}
\def\@versim#1#2{\lower 2.9truept \vbox{\baselineskip 0pt \lineskip
    0.5truept \ialign{$\m@th#1\hfil##\hfil$\crcr#2\crcr\sim\crcr}}}
\catcode`\@=12

\def\eps{{\cal E}}

\def\ln{\hbox{${\rm ln}\, $}}

\def\pr{p_{\rm r}}
\def\pt{p_{\rm t}}

\def\psit{\Psi_{\rm T}}

\def\ra{r_{\rm a}}
\def\rai{r_{\rm ai}}

\def\sigr{\sigma_{\rm r}}
\def\sigt{\sigma_{\rm t}}

   \title[On the global density slope--anisotropy inequality]
   {How general is the global density slope--anisotropy inequality?}

   \author[Ciotti \& Morganti]
           {Luca Ciotti \& Lucia Morganti\thanks{Current address:
 Max-Planck-Institut f\"ur Ex. Physik, Giessenbachstra\ss{}e,
 D-85741 Garching, Germany}\\Astronomy Department, 
      University of Bologna, via Ranzani 1, 40127 Bologna, Italy}

\date{Accepted version, June 11, 2010}
\pubyear{2010}

\begin{document}
 \maketitle

\begin{abstract} 
  Following the seminal result of An \& Evans, known as the central
  density slope--anisotropy theorem, successive investigations
  unexpectedly revealed that the density slope--anisotropy inequality
  holds not only at the center, but at all radii in a very large class
  of spherical systems whenever the phase--space distribution function
  is positive.  In this paper we derive a criterion that holds for all
  spherical systems in which the augmented density is a separable
  function of radius and potential: this new finding allows to unify
  all the previous results in a very elegant way, and opens the way
  for more general investigations.  As a first application, we prove
  that the global density slope--anisotropy inequality is also
  satisfied by all the explored additional families of
  multi--component stellar systems.  The present results, and the
  absence of known counter--examples, lead us to conjecture that the
  global density slope--anisotropy inequality could actually be a
  universal property of spherical systems with positive distribution
  function.
\end{abstract}

\begin{keywords}
celestial mechanics -- stellar dynamics -- galaxies: kinematics and dynamics
\end{keywords}

\section{Introduction}
In the study of stellar systems based on the ``$\rho$--to--$f$''
approach (where $\rho$ is the density distribution and $f$ is the
associated phase--space distribution function, hereafter DF), $\rho$
is given, and specific assumptions on the internal dynamics of the
model are made (e.g. see Bertin 2000; Binney \& Tremaine 2008,
hereafter BT08).  In some special cases inversion formulae exist and
the DF can be recovered in integral form or as series expansion (see,
e.g., Fricke 1952; Lynden--Bell 1962; Osipkov 1979; Merritt 1985;
Dejonghe 1986, 1987; Cuddeford 1991; Hunter \& Qian 1993; Ciotti \&
Bertin 2005).  Once the DF of the system is known, a non--negativity
check should be performed, and in case of failure the model must be
discarded as unphysical, even if it provides a satisfactory
description of data.  Indeed, a minimal but essential requirement to
be met by the DF (of each component) of a stellar dynamical model is
positivity over the accessible phase--space.  This requirement, the
so--called \textit{phase--space consistency}, is much weaker than the
model stability, but it is stronger than the fact that the Jeans
equations have a physically acceptable solution.  However, the
difficulties inherent in the operation of recovering analytically the
DF prevent in general a simple consistency analysis.

Fortunately, in special circumstances phase--space consistency can be
investigated without an explicit recovery of the DF.  For example,
analytical necessary and sufficient conditions for consistency of
spherically symmetric multi--component systems with Osipkov--Merritt
anisotropy (Osipkov 1979, Merritt 1985, hereafter OM) were derived in
Ciotti \& Pellegrini (1992, hereafter CP92; see also Tremaine et
al. 1994) and applied in several investigations (e.g., Ciotti 1996,
1999; Ciotti \& Lanzoni 1997; Ciotti \& Morganti 2009, hereafter CM09;
Ciotti, Morganti \& de Zeeuw 2009).  Moreover, in Ciotti \& Morganti
(2010, hereafter CM10a) analytical necessary and sufficient
consistency criteria for the family of spherically symmetric,
multi--component, generalized Cuddeford systems (which contain as very
special cases constant anisotropy and OM systems) have been derived.

Another necessary condition for consistency, the focus of this paper,
is the ``central cusp--anisotropy theorem'' (An \& Evans 2006,
hereafter AE06; see also equation~[28] in de Bruijne et al. 1996), an
inequality relating the values of the central density slope $\gamma$
and of the central anisotropy parameter $\beta$ of consistent
spherical systems, namely $\gamma\geq2\beta$ (see Section 2).  This
condition was derived for constant anisotropy systems, and then
generalized asymptotically to the central regions of spherical systems
with arbitrary anisotropy distribution.  In AE06 it was also shown
that the density slope--anisotropy inequality actually holds
rigorously at \textit{every} radius in constant anisotropy systems,
and not only at their center.  We will refer to this case, i.e. when
$\gamma(r)\geq2\beta(r)\quad\forall r$, as the \textit{Global
  Density--Slope Anisotropy Inequality} (hereafter GDSAI).

Surprisingly, in CM09 we showed that the CP92 necessary condition for
model consistency is nothing else than the GDSAI in disguise.  Thus,
not only in constant anisotropy systems but also in each component of
multi--component OM systems the GDSAI holds.  Prompted by this curious
result, in CM10a we introduced the larger family of multi--component
generalized Cuddeford systems, we studied their phase--space
consistency, and we finally proved that the GDSAI is again a necessary
condition for phase--space consistency of each density component.

The results of CM09 and CM10a revealed an unexpected generality of the
GDSAI and, in absence of known counter--examples (see also the
discussion in CM10a), it is natural to ask whether the GDSAI is even
more general, i.e. it is necessarily obeyed by all spherically
symmetric, two--integrals systems with positive DF.  If such
conjecture proved true, it would be remarkable not only from a
theoretical point of view, but also for applications. In fact, as it
would hold separately for each density component of a stellar system,
the value of the anisotropy parameter of the stellar component would
be controlled at each radius by the local stellar density slope
itself, independently of the dark matter halo.  This constraint could
be then used to reduce the impact of mass--anisotropy degeneracy in
observational works\footnote{Clearly, as by definition $\beta\leq1$,
  such limitation would be useful only in the regions where the
  density slope is $\leq 2$, while in the galaxy outskirts the
  inequality is not helpful, as $\gamma>2$ for mass convergence.}.
Motivated by the above arguments, we searched for a proof of the
general validity of the GDSAI.  While a proof is still missing, some
relevant advance has been made: in particular, we obtained a new
criterion that allows us not only to prove in a very elegant and
unified way all our previous results, but also to demonstrate that new
families of models do necessarily obey the GDSAI when consistent.

The paper is organized as follows.  In Section 2 a general criterion
linking phase--space consistency to the GDSAI for systems whose
augmented density is a separable function of radius and potential is
derived, which unifies all the results obtained so far on the GDSAI,
and opens the way to the investigation of an even wider class of
models.  In Section 3 the new criterion is used to prove in a new way
that the GDSAI is obeyed by generalized Cuddeford models, but also by
some well-known stellar systems not belonging to the family of
generalized Cuddeford models, thus further extending the validity of
the GDSAI as a necessary condition for phase--space consistency.
Finally, the main conclusions are summarized in Section4.

\section{A general criterion}
In Ciotti \& Morganti (2010, hereafter CM10b) we showed analytically,
by direct computation, that two well-known anisotropic models not
belonging to the generalized Cuddeford family, namely the Dejonghe
(1987) anisotropic Plummer model, and the Baes \& Dejonghe (2002)
anisotropic Hernquist model, indeed obey the GDSAI whenever their DF
is positive.  On one hand this result proves that the global
inequality is not a specific property of generalized Cuddeford models,
due perhaps to some special dependence of their DFs on energy and
angular momentum.  On the other hand, it reinforces the conjecture
that the GDSAI could be a very general (if not a universal) property
of consistent spherical models.  In this Section we provide a new hint
to the latter hypothesis, as we show, with the aid of a new criterion,
how the GDSAI is rigorously valid in a very large class of consistent
systems, containing not only the multi--component generalized
Cuddeford systems and the two models discussed in CM10b, but whole new
families of models.

\subsection{General relations and the case of factorized systems}
We consider stationary, non--rotating, spherically symmetric systems
with a two--integrals phase--space distribution function $f=f(\eps,
J)$, where $\eps=\psit-v^2/2$ is the relative energy per unit mass,
$\psit=-\Phi_{\rm T}$ is the relative total potential, and $J$ is the
angular momentum modulus per unit mass.  In general $\psit$ may
contain also the contribution of an ``external'' potential (for
instance the one corresponding to a dark matter halo).

It is easy to show (e.g. see BT08, Ciotti 2000)
that the density distribution $\rho$,
the radial velocity dispersion $\sigr$,
and the tangential velocity dispersion $\sigt$
are related to the DF as
\begin{equation}\label{rho}
\rho=\frac{4\pi}{r^2}\int_0^{\psit}d\eps
\int_0^{J_m}\frac{f(\eps, J)J}{\Delta}dJ,
\end{equation}
\begin{equation}\label{sigmarad}
\rho\sigr^2=\frac{4\pi}{r^2}\int_0^{\psit}d\eps
\int_0^{J_m}f(\eps, J)\Delta JdJ=\pr,
\end{equation}
\begin{equation}\label{sigmatan}
\rho\sigt^2=\frac{4\pi}{r^4}\int_0^{\psit}d\eps
\int_0^{J_m}\frac{f(\eps, J)J^3}{\Delta}dJ=\pt,
\end{equation}
where $\Delta=\sqrt{2(\psit-\eps)-J^2/r^2}$ and
$J_m=r\sqrt{2(\psit-\eps)}$.  With $\pr$ and $\pt$ we indicate the
radial and tangential pressure, respectively.

Given the identities above, and given the definition of the anisotropy
parameter $\beta$, easy algebra proves the remarkable identities
\begin{eqnarray}\label{pressure}
&&\rho(r,\psit)=\frac{\partial\pr}{\partial\psit}, \\
&&\beta(r,\psit)\equiv 1-\frac{\pt}{2\pr}=-\frac{1}{2}\frac{\partial\ln\pr}{\partial\ln r}
\end{eqnarray}
(e.g., see Spies \& Nelson 1974, Dejonghe 1986, Cavenago 1987,
Dejonghe \& Merritt 1992, Bertin et al. 1994, Bertin 2000, Baes \& van
Hese 2007).  Note that these relations hold independently of the
specific radial dependence of $\psit$, and that the radial trends of
$\rho$ and $\beta$ are known only after the total potential $\psit$ is
given.  The existence of the general relations~(\ref{pressure})
and~(5) is of the greatest importance for the present study, because
it shows clearly that $\rho$ (and so $\gamma$) and $\beta$ are
somewhat linked by the function $\pr$.  This link could open the way
to a general proof of the GDSAI.

At the present stage, we do not attempt a general proof, but we focus
our attention on a special case, that of spherical systems with a
factorized augmented density.  In practice, we consider special
spherical systems in which the radial pressure $\pr$ is a factorized
function of the radius and of the total potential.  Therefore, from
equation~(\ref{pressure}), it follows that also the density
distribution can be written as
\begin{equation}\label{rhofac}
\rho(r,\psit)\equiv A(r) B(\psit);
\end{equation}
of course, while the function $B$ in the expression above is the
derivative of the potential dependent factor in the factorized
expression of $\pr$, the radial function $A$ is the same.  We note
that the multi--component generalized Cuddeford models (and therefore
also the constant anisotropy, OM, and Cuddeford models, see CM10a), as
well as the two models discussed in CM10b, belong to such family of
systems.  From equation~(5) it follows that, independently of the
specific radial dependence of $\psit$,
\begin{equation}\label{betafac}
 2\beta(r)=-\frac{d\ln A}{d\ln r}.
\end{equation}

Now, for assigned $\psit(r)$, equation~(\ref{pressure}) shows that
the logarithmic slope of the density profile~(\ref{rhofac}) 
can be written as
\begin{equation}\label{gammafac}
\gamma(r)\equiv-\frac{d\ln\rho}{d\ln r}=
-\frac{d\ln B}{d\ln\psit}\frac{d\ln\psit}{d\ln r}+2\beta(r);
\end{equation}
since from Newton theorem $\psit(r)$ is a monotonically decreasing
function of radius, identity~(\ref{gammafac}) proves the following

\textbf{Criterion~1}: in all spherical systems
whose density distribution is a separable function of radius and total
potential, $\rho=A(r) B(\psit)$, the global inequality
$\gamma(r)\geq2\beta(r)\quad\forall r$ holds $\Leftrightarrow
B(\psit)$ is a monotonically increasing function of $\psit$.  \medskip

Therefore, if one is able to show that in all factorized consistent
systems the $B$ function is necessarily monotonic, then the GDSAI will
hold in all these systems.  We were not able to prove or disprove this
possibility in general, however in Section~3 we present interesting
results along this line.

Before moving to discuss the new results, we note the following
curious fact: the GDSAI can also be expressed as a condition on the
radial velocity dispersion of the model.  In fact, for a
two--integrals spherical system, the relevant Jeans equation can be
written as
\begin{equation}\label{jeans}
\frac{d\rho\sigr^2}{dr}+\frac{2\beta\rho\sigr^2}{r}=\rho\frac{d\psit}{dr}
\end{equation}
(e.g., BT08).
Introducing the logarithmic density slope
as in equation~(\ref{gammafac}), and rearranging the terms,
one finds
\begin{equation}\label{jeansGamma}
\gamma(r)-2\beta(r)=r\left(\frac{d\sigr^2}{dr}-\frac{d\psit}{dr}\right)\geq0
\end{equation}
as an equivalent, alternative formulation of the GDSAI.
Unfortunately, despite the deceptively simple form,
for a given family of consistent models
proving the necessity of the GDSAI from equation~(\ref{jeansGamma}) 
is not easier than working directly on phase--space.

\section{Results}
In this Section we apply Criterion~1 to two families of models, and we
show that also these systems not belonging to the generalized
Cuddeford family obey the GDSAI in case of phase--space consistency.
However, in order to illustrate the power of the new result, with the
aid of Criterion~1 we first re--derive almost immediately the results
obtained in CM10a with some lengthy algebra for the generalized
Cuddeford models.

\subsection{A new proof of the GDSAI
for generalized Cuddeford systems}
As described in CM10a, the DF of each component 
of generalized Cuddeford systems
is described by the sum of an arbitrary number 
of Cuddeford (1991) DFs with arbitrary positive weights $w_i$,
\begin{equation}\label{sumCud}
f=J^{2\alpha}\sum_i w_i h(Q_i),\quad Q_i=\eps-\frac{J^2}{2\rai^2},
\end{equation}
and possibly different anisotropy radii $\rai$, but same $h$ function
and angular momentum exponent $\alpha$.  Here, $\alpha$ is a real
number $>-1$, and $h(Q_i)=0$ for $Q_i\leq0$.  Note that
multi--component OM models (CP92, CM09), and constant anisotropy
models are special cases of equation~(\ref{sumCud}), obtained for
$i=1$ and $\alpha=0$, and $i=1$ and $\ra=\infty$, respectively.

As shown in CM10a, the spatial density associated with the DF~(\ref{sumCud}) 
is a factorized function as in equation~(\ref{rhofac}), where
\begin{eqnarray}\label{aCudsum}
&&A(r)=(2\pi)^{3/2}\frac{\Gamma(\alpha+1)}{\Gamma(\alpha+3/2)}\sum_i\frac{ w_i
r^{2\alpha}}{(1+r^2/\rai^2)^{\alpha+1}},\\
&&B(\psit)=\int_0^{\psit}(\psit-Q)^{\alpha+1/2}h(Q)dQ,
\end{eqnarray}
and $\Gamma(x)$ is the complete gamma function.  Therefore, the
anisotropy parameter $\beta$, derived in CM10a by direct computation,
can now be obtained from equations~(\ref{aCudsum})
and~(\ref{betafac}), and Criterion~1 can be applied to the family of
generalized Cuddeford models.  In CM10a it was shown with some algebra
(see equations~[19] and~[25] therein) that for $\alpha\geq-1/2$ the
first of the necessary conditions for phase--space consistency can be
rewritten as the GDSAI.  A more elegant proof can now be obtained by
using Criterion~1.  In fact, direct differentiation of equation~(23)
shows that $B$ is a monotonically increasing function of $\psit$
whenever $h>0$ and $\alpha\geq-1/2$.  An interesting situation arises
for $\alpha=-1/2$.  In this case equation~(28) in CM10a shows that the
GDSAI is both necessary and sufficient condition for phase--space
consistency: in the present context $d B/d\psit=h(\psit)$, and so the
monotonicity of $B$ is ensured if and only if $h$ is positive,
confirming the CM10a result.  We are left with the case
$-1<\alpha<-1/2$.  As shown in CM10a, in this case only a sufficient
condition, coincident with the~GDSAI, is available.  What are the
information that can be derived from Criterion~1 in this case?
Indeed, for $-1<\alpha<-1/2$, it is not possible to compute the
derivative of equation~(23) directly, and so to test the monotonicity
of $B$.  However, this problem can be circumvented by first
integrating by parts, and then performing differentiation, obtaining
\begin{equation}\label{dpsiCud}
 \frac{d B}{d\psit}=\psit^{\alpha+1/2} h(0)+\int_{0}^{\psit} (\psit-Q)^{\alpha+1/2}h'(Q)dQ.
\end{equation}
From the expression above, we conclude that the GDSAI is again
necessary for phase--space consistency in all generalized
multi--component Cuddeford systems with $-1<\alpha<-1/2$ and
$h'(Q)>0$.

Summarizing, for $\alpha>-1/2$ the GDSAI is necessary for consistency,
for $\alpha=-1/2$ it is equivalent (i.e., necessary and sufficient),
and for $-1<\alpha<-1/2$ it is just sufficient, but also necessary if
$h'(Q)>0$.  Of course, whenever the last condition is not satisfied,
the possibility to build a counter--example to the conjecture that the
GDSAI is a universal \textit{necessary} condition for consistency
remains open.  If one could prove that for $-1<\alpha<-1/2$ the
inequality is necessary whenever $h>0$, then one would conclude that
in multi--component generalized Cuddeford models the GDSAI is
equivalent to consistency for $-1<\alpha\leq-1/2$.

\subsection{The Baes \& van Hese (2007) anisotropic models}
By using Criterion~1 we now show
that the GDSAI is obeyed by another (quite large) family of model.

Baes \& van Hese (2007, Section 4.2) considered 
the family of augmented density profiles
\begin{equation}\label{rhobavanhese}
\rho=\rho_0 \left(\frac{r}{\ra}\right)^{-2\beta_0} \left(1+\frac{r^{2\delta}}{\ra^{2\delta}}\right)^{\frac{\beta_0-\beta_\infty}{\delta}} \left(\frac{\Psi}{\Psi_0}\right)^p\left(1-\frac{\Psi^s}{\Psi_0^s}\right)^q,
\end{equation}
where $\delta>0$, $q\leq0$, $s>0$, $0\leq\Psi\leq\Psi_0$,
and $\Psi_0$ is the value of the central relative potential.
From equation~(\ref{betafac}) the anisotropy profile is
\begin{equation}
\beta(r)=\frac{\beta_0+\beta_\infty(r/\ra)^{2\delta}}{1+(r/\ra)^{2\delta}},
\end{equation}
so that $\beta_0$ and $\beta_\infty$ are the values of the orbital
anisotropy parameter at small and large radii, respectively; since the
anisotropy parameter $\beta$ cannot exceed the value of $1$ in case of
positive DF, both $\beta_0$ and $\beta_\infty$ are $\leq1$.

Note that the requirement of finite total mass for the density
distribution~(\ref{rhobavanhese}), i.e. $\Psi(r)\sim1/r$ for
$r\to\infty$, translates into the condition $p+2\beta_\infty>3$, and
from the limitation on $\beta_\infty$ it follows that $p\geq1$.  Now,
it is easy to show that $d B(\Psi)/d\Psi>0$ for $0\leq\Psi\leq\Psi_0$,
when $s>0$, $q\leq0$, and $p\geq1$, and thus Criterion~1 ensures that
the GDSAI is necessarily obeyed also by these profiles.  In the
particular case $q=0$, it is possible to prove again the GDSAI by
using equation~(\ref{jeansGamma}) and considering the expression of
the radial velocity dispersion given in equation~(37) of Baes \& van
Hese (2007).  Note also that the analytical DF associated to these
systems is
\begin{equation}
f=\sum_k\eps^{p+ks-3/2}g_k\left(\frac{J^2}{2\eps}\right),
\end{equation}
where the $g_k$ are hypergeometric functions
(see equations~[41] and [42] in Baes \& van Hese 2007),
and so in this case the DF 
is not of the Cuddeford generalized family.

We finally note that the two models discussed analytically in CM10b,
i.e. the Dejonghe~(1987) anisotropic Plummer model, and the Baes \&
Dejonghe~(2002) anisotropic Hernquist model, are both special cases of
the family~(\ref{rhobavanhese}), and therefore the fact that they also
obey the GDSAI is just a special case of the result of this Section.

\subsection{The Cuddeford \& Louis (1995) anisotropic polytropes}
We finally consider the family of models introduced by Cuddeford \&
Louis (1995).  At variance with the previous models, these systems are
not introduced by using the augmented density technique, but from
their DF, in a way similar to the models in Section~3.1, so that
Criterion~1 cannot be applied directly.  Their DF is
\begin{equation}\label{dfCudLouis}
f(\eps,J) = \eps^{q-2}h(k), \qquad k\equiv\frac{ J^2}{2\ra^2\eps},
\end{equation}
where $\eps\geq0$, and $\ra$ is the anisotropy radius.
These models are known as \textit{anisotropic polytropes},
and the special case $h(k)=(1+k)^\alpha$
was studied by Louis (1993).
The formulae for $\rho$ and $\pr$
were already obtained also for the more general case of $f=g(\eps)h(k)$,
and here we just report the result for the case~(\ref{dfCudLouis}):
%
%
\begin{eqnarray}\label{gen}
&&\rho=2^{3/2}\pi{\rm{B}}(q,1/2)\psit^{q-1/2}\eta^{q-1}\int_0^\infty\frac{h(k)dk}{(k+\eta)^q},\\
&&\pr=2^{5/2}\pi{\rm{B}}(q,3/2)\psit^{q+1/2}\eta^{q-1}\int_0^\infty\frac{h(k)dk}{(k+\eta)^q},
\end{eqnarray}
where $\rm{B}$ is the complete Beta function and $\eta=r^2/\ra^2$.
Incidentally, the validity of equations~(\ref{gen}) and~(20) can be
easily checked by using equation~(\ref{pressure}).  Note that the
convergence of the energy integral in equation~(\ref{gen}) requires
$q>0$ near $\eps=0$.  As $\pr$ and $\rho$ are in factorized form (see
also Dejonghe 1986, Sect. 1.7.3), we can apply Criterion~1: since
$\rho\propto\psit^{q-1/2}$, the GDSAI is satisfied whenever
$q\geq1/2$.  This result can be obtained immediately also from
equation~(\ref{jeansGamma}), as the radial velocity dispersion has the
remarkably simple expression
\begin{equation}\label{siglc}
\sigr^2=\frac{\pr}{\rho}=\frac{2\psit}{2q+1}.
\end{equation}
The situation is less straightforward in the interval $0<q<1/2$.  In
fact, if a consistent model exists in this interval, then it will
represent a case of violation of the GDSAI.  Therefore, it is natural
to ask whether it is possible to construct a consistent dynamical
model with $q>0$, but violating the GDSAI ($q<1/2$).  We are not able
to answer this question in general, however we note that in the
self--consistent case of finite total mass in which $\psit\sim1/r$ for
$r\to\infty$, volume integration of equation~(\ref{gen}) and
successive inversion of order of integration shows that $q>3/2$ is
necessary in order to have a finite mass for the component under
scrutiny (with possible further restrictions dependent on the
asympotic nature of the $h$ function).  In other words, there are not
consistent anisotropic polytropes with finite mass and $q<3/2$.
Therefore the GDSAI is a necessary condition for anisotropic
polytropes in generic external potential when $q\geq1/2$, and for all
finite mass self--consistent models.  It remains open the possibility
of existence of consistent anisotropic polytropes (of infinite mass)
with $0<q<1/2$, which would violate the GDSAI.

Actually the discussion above can be generalized, similarly to what we
did in CM10a when we constructed the family of generalized Cuddeford
anisotropic systems.  In fact, we now consider the family of
generalized anisotropic polytropes with DF
\begin{equation}\label{dfgenCudLouis}
f(\eps,J) = \eps^{q-2}\sum_i w_i h(k_i), \qquad k_i\equiv\frac{ J^2}{2\rai^2\eps},
\end{equation}
with different anisotropy radii $\rai$
and positive weights $w_i$.
With the definition $\eta_i=r^2/\rai^2$,
equations~(\ref{gen}) and~(20) become
\begin{eqnarray}\label{gengen}
&&\rho=2^{3/2}\pi{\rm{B}}(q,1/2)\Psi^{q-1/2}\sum_iw_i\eta_i^{q-1}\int_0^\infty\frac{h(k)dk}{(k+\eta_i)^q},\\
&&\pr=2^{5/2}\pi{\rm{B}}(q,3/2)\Psi^{q+1/2}\sum_iw_i\eta_i^{q-1}\int_0^\infty\frac{h(k)dk}{(k+\eta_i)^q},
\end{eqnarray}
and it is immediate to see that all the previous conclusions
hold also for these general models,
even though their DF is not of the family~(\ref{dfCudLouis}).

\section{Discussion and conclusions}

After the discovery of the ``central cusp--anisotropy theorem''
(AE06), an inequality relating the central value of the density slope
and the anisotropy parameter in consistent stellar systems, successive
investigations (CM09, CM10a) unexpectedly revealed that the density
slope--anisotropy inequality holds not only at the center, but at all
radii in a very large class of spherical systems (the generalized
multi--component Cuddeford systems), whenever the phase--space
distribution function is positive. We call this latter inequality the
\textit{Global Density Slope Anisotropy Inequality} (GDSAI).

In absence of known counter--examples, i.e. two--integrals stellar
systems with positive DF but violating the GDSAI, in this paper we
focused on the possibility that the GDSAI is actually universal,
i.e. it is necessarily obeyed by all spherically symmetric,
two--integrals systems with positive DF.  If such conjecture proved
true, it would be remarkable not only from a theoretical point of
view, but also for applications.  In fact, it could be used to reduce
the impact of mass--anisotropy degeneracy in observational works, as
orbital anisotropy would be in some sense controlled by the local
density slope of the stellar distribution in galaxies (in the inner
regions where $\gamma\leq2$).  While a proof of this conjecture is
still missing, some relevant advance has been made: in particular, we
obtained a new criterion that allows us not only to prove in a simple,
very elegant, and unified way all the previously known results, but
also to investigate new families of multi--component models.  The main
results of this paper can be summarized as follows:
\begin{enumerate}
\item By using two previously known and fully general identities
  relating the density profile, the anisotropy profile, and the radial
  pressure in two--integrals systems, specialized to the case of
  factorized systems, we found a very simple condition equivalent to
  the GDSAI, namely that the potential dependent function in the
  augmented density is monotonically increasing.
\item As a first application of the new condition, we showed that all
  the previous cases, each of them proved with ``ad-hoc'' analysis,
  are in fact all simple cases of the new relation, and the proof is
  almost immediate.
\item The new criterion is then applied to extend the validity of the
  GDSAI to other models, namely the Baes \& van Hese (2007)
  anisotropic models, and the Cuddeford \& Louis (1995) anisotropic
  polytropes.  This latter family is extended to generalized
  multi--component anisotropic polytropes, and it is shown that also
  in this case the GDSAI holds.
\item As we are not able to show that the monotonicity of the
  potential dependent function in the factorized augmented density is
  necessarily monotonic whenever the DF is positive, our investigation
  leaves open the possibility that some spherical systems with
  positive phase--space distribution function may violate the GDSAI.
  Such examples, if they exist, may be found (in the class of models
  studied so far) only in the family of Cuddeford models with angular
  momentum exponent in the range $-1<\alpha<-1/2$, or in the family of
  Cuddeford \& Louis (1995) anisotropic polytropes with infinite mass
  and $0<q<1/2$.
\end{enumerate}

Finally, we conclude by noticing that the proof of the general
validity of the GDSAI could be obtained by using
equations~(\ref{pressure}) and~(5) in all their generality.  At the
present stage we do not have such a proof; however, neither
counter--exemples are known to us.  Moreover, we note that supporting
arguments are provided by numerical simulations of N--body systems,
whose end--products show correlations between $\beta$ and $\gamma$
(e.g., see Hansen \& Moore 2006, Mamon et al. 2006).  In addition,
Michele Trenti kindly provided us with a large set of numerically
computed $f_{\nu}$ models (Bertin \& Trenti 2003) and all of them,
without exception, satisfy the GDSAI.  We also verified numerically
that the GDSAI is satisfied by a large set of radially anisotropic
Hernquist and Jaffe models with quasi--separable DF constructed by
Gerhard (1991) by using the so--called $h_{\alpha}$ circularity
function.

\section*{Acknowledgments} 
The authors wish to thank the referee Maarten Baes for a careful
reading of the paper and for comments that improved the presentation.
L.C. thanks the Princeton Institute for Computational Science and
Engineering (PICSciE), and the Department of Astrophysical Sciences of
Princeton University where part of this work was done.  L.M. thanks
Ortwin Gerhard for providing the numerical code used to generate
anisotropic models by using the circularity function.


\end{document}